\documentclass[review]{elsarticle}

\usepackage{lineno,hyperref}
\modulolinenumbers[5]

\usepackage{graphicx}
\usepackage{amsmath}
\usepackage{amssymb}
\usepackage{color}
\usepackage[normalem]{ulem}

\begin{document}

\begin{frontmatter}

\title{Breather dynamics in a stochastic sine-Gordon equation: evidence of noise-enhanced stability}


\author[unipa]{Duilio De Santis\corref{correspondingauthor}}
\cortext[correspondingauthor]{Corresponding author}
\ead{duilio.desantis@unipa.it}
\author[unisa]{Claudio Guarcello}
\author[unipa,unn]{Bernardo Spagnolo}
\author[unipa]{Angelo Carollo}
\author[unipa]{Davide Valenti}
\address[unipa]{Dipartimento di Fisica e Chimica ``E.~Segr\`{e}", Group of Interdisciplinary Theoretical Physics, Università degli Studi di 
Palermo, I-90128 Palermo, Italy}
\address[unisa]{Dipartimento di Fisica ``E.~R.~Caianiello", Università degli Studi di Salerno, I-84084 Fisciano, Salerno, Italy}
\address[unn]{Radiophysics Department, Lobachevsky State University, 603950 Nizhniy Novgorod, Russia}

\begin{abstract}
The dynamics of sine-Gordon breathers is studied in the presence of dissipative and stochastic perturbations. Taking a stationary breather with a random phase value as the initial state, the performed simulations demonstrate that a spatially-homogeneous noisy source can make the oscillatory excitation more stable, i.e., it enables the latter to last significantly longer than it would in a noise-free scenario. Both the frequency domain and the localization of energy are examined to document the effectiveness of the noise-enhanced stability phenomenon, which emerges as a nonmonotonic behavior of an average characteristic time for the breather as a function of the noise intensity. The influence of the mode's starting frequency on the results and their robustness against an additional thermal background are also addressed.
\end{abstract}


\end{frontmatter}

\section{Introduction}
\label{sec1}

Solitons, i.e., localized nonlinear waves with prominent particle-like behavior, represent one of the most spectacular fingerprints of nonlinearity in nature~\cite{Scott_2006}. Mathematically, soliton solutions are found in several completely integrable (nonlinear) partial differential equations, e.g., the Korteweg-de Vries, the nonlinear Schr\"{o}dinger, and the sine-Gordon (SG) equation~\cite{Ablowitz_1981}. These systems catch, in a sense, the core features of many processes involving nonlinearity, thus they emerge as approximate models for a large variety of phenomena~\cite{Scott_2003, Dauxois_2006, Remoissenet_2013}. For most real-world purposes, integrability-breaking perturbation terms have to be included to account for, e.g., external forces, impurities, dissipation, and environmental coupling~\cite{Kivshar_1989}. However, being outstandingly robust, solitons (not to be intended in the strict mathematical sense anymore) are able to preserve their main features under such harsh circumstances. As a result, the soliton theory has led to all kinds of striking experimental observations and applications over the years~\cite{Scott_2003, Dauxois_2006, Remoissenet_2013}.

In any system that admits solitonic excitations, both properly understanding their dynamics and mastering generation techniques constitute crucial problems that are not easily solvable. Therefore, significant research efforts have been devoted to these tasks in different contexts, the most common investigation tools being perturbative analyses and numerical simulations~\cite{Kivshar_1989}.

The present paper is focused on the SG equation~\cite{Jesus_2014}, i.e., the classical wave equation with a nonlinear sine source term. Such a mathematical framework finds application in pendula, Josephson systems, gravity, high-energy physics, biophysics, and seismology~\cite{Jesus_2014, Ivancevic_2013, Yakushevich_2021, Bykov_2014}. Both topologically charged solitonic solutions, i.e., kinks and antikinks, and neutral, i.e., breathers, are of fundamental importance in the SG system. In this regard, an intriguing open challenge is the experimental detection of breathers in long Josephson junctions (LJJs)~\cite{Barone_1982}. Indeed, while kinks (or fluxons, since they represent magnetic flux quanta in LJJs) are readily trackable~\cite{Parmentier_1993, Ustinov_1998} and can nowdays be exploited in various ways~\cite{Soloviev_2014, Soloviev_2015, Guarcello_2017, Gua18, GuaSol18, GuaSolBra18, Gua19, Wustmann_2020, Dobrowolski_2020, Osborn_2021, Gatlik_2021}, breathers (i.e., kink-antikink oscillatory bound states) have yet to be observed, despite plenty of theoretical works on the subject~\cite{McLaughlin_1978_PRA, Lomdahl_1986, Jensen_1991, Jensen_1992, Gulevich_2006, Gulevich_2012, Johnson_2013, De_Santis_2022, De_Santis_2022_arxiv}. The main problems connected with the breathers' detection are that~\cite{McLaughlin_1978_PRA, Gulevich_2006, Gulevich_2012} (i) time-averaged voltage measurements cannot probe them due to their rapid pulsations, and (ii) dissipation leads to their radiative decay.

As stated by Guvelich \textit{et al.}~\cite{Gulevich_2012}, for a breather detection experiment to succeed, the breather mode should be first efficiently excited, and then maintained in a given area to carry out the measurements. Here, the second task is faced, see Refs.~\cite{De_Santis_2022, De_Santis_2022_arxiv} for recent discussions concerning generation techniques. More precisely, the manuscript's main scope is to understand whether stochastic perturbations can positively impact the breather's dynamics in the presence of dissipation. In fact, while previous literature~\cite{Lomdahl_1986, Jensen_1991, Jensen_1992} has established that, for specific breather initial configurations, a monochromatic force can compensate for the radiative losses and stabilize the mode, the effects of noise sources have remained relatively unexplored. Since disorder may come into different forms in a Josephson device~\cite{Castellano_1996, Fedorov_2007, Fedorov_2008, Fedorov_2009, Valenti_2014, Guarcello_2015_1, Guarcello_2016, Guarcello_2017_Graphene}, with significant consequences on its response, such a topic is relevant for identifying an optimal setup to observe the breather state, but also for all its subsequent applications, e.g., in information transmission~\cite{Macias-Diaz_2007, Macias-Diaz_2007_1} and quantum computation~\cite{Fujii_2007, Fujii_2008, Fuj09}.

To investigate the influence of fluctuations on this soliton's dissipative dynamics, the perturbed SG equation is numerically integrated, beginning from a stationary breather with a random initial phase value. This condition is chosen here with the intention of mimicking the successful generation of the breather mode in a confined area, even without precise control of its internal parameters. Interestingly, the simulations indicate that a noisy force uniformly distributed in space (i.e., an electrical current in the case of an LJJ) can sustain the excitation far beyond its radiative decay lifetime. In other words, the window of availability for measurements can be stochastically increased. This is documented by looking at both the frequency domain and the localization of energy. Furthermore, a nonmonotonic behavior of an average characteristic time for the breather as a function of the noise intensity is found, which is a manifestation of the so-called noise-enhanced stability effect~\cite{Valenti_2014, Mantegna_1996, Agudov_2001, Spagnolo_2004, Pankratov_2004, Fiasconaro_2005, D'Odorico_2005, Hurtado_2006, Spagnolo_2007, Mankin_2008, Yoshimoto_2008, Fiasconaro_2009, Trapanese_2009, Fiasconaro_2010, Li_2010, Smirnov_2010, Shongwe_2015, Guarcello_2015_Graphene, Guarcello_2021}. The dependence of the results on the breather's starting frequency and their robustness against an additional thermal background are also examined.

Before proceeding any further, it should be noted that, although this study originates within the LJJ realm, the discussion is kept on fairly general grounds, and the results are not restricted to any specific domain of application of the model. Indeed, SG breathers have also been recently studied in cuprate superconductors~\cite{Dienst_2013}, geology~\cite{Zalohar_2020}, and DNA systems~\cite{Liu_2021}, to name but a few.

The paper is structured as follows. Section~\ref{sec2} defines the physical model. Section~\ref{sec3} presents the obtained results for an increasingly complex perturbed SG equation. More explicitly, an overview of the (noise-free) dissipative case is given in Sec.~\ref{sec3a}, while Sec.~\ref{sec3b} and Sec.~\ref{sec3c} are concerned with, respectively, the effects of a spatially-homogeneous noisy force and the robustness of the findings with respect to an additional thermal background. Finally, conclusions are drawn in Sec.~\ref{sec4}.

\section{The model}
\label{sec2}

For the field ${ \varphi(x, t) }$, the following perturbed SG equation is considered
\begin{equation}
\label{eqn:1}
\varphi_{xx} - \varphi_{tt} - \alpha \varphi_{t} = \sin \varphi - \gamma(t) - \gamma_T (x, t)
\end{equation}
in the spatio-temporal domain ${ -l/2 \leq x \leq l/2 }$ and ${ t \geq 0 }$. Here, a subscript notation has been used to denote partial differentiation, i.e., ${ \partial \varphi / \partial t = \varphi_t }$, $ \alpha $ is a damping parameter, ${ \gamma(t) }$ is a spatially-uniform Gaussian noisy force with
\begin{equation}
\label{eqn:2}
\langle \gamma(t) \rangle = 0 \text{\phantom{0}and\phantom{0}} \langle \gamma(t_1) \gamma(t_2) \rangle = 2 \Gamma \delta (t_1 - t_2) ,
\end{equation}
and ${ \gamma_T (x, t) }$ is a delta-correlated Gaussian thermal background
\begin{equation}
\label{eqn:3}
\langle \gamma_{T}(x, t) \rangle = 0 \text{\phantom{0}and\phantom{0}} \langle \gamma_{T}(x_1, t_1) \gamma_{T}(x_2, t_2) \rangle = 2 \alpha \Gamma_T \delta (x_1 - x_2) \delta (t_1 - t_2) .
\end{equation}
No-flux boundary conditions are imposed to solve Eq.~\eqref{eqn:1}
\begin{equation}
\label{eqn:4}
\varphi_x (-l/2, t) = \varphi_x (l/2, t) = 0 ,
\end{equation}
along with the initial conditions
\begin{equation}
\label{eqn:5}
\varphi (x, 0) = \varphi^b (x, 0) \text{\phantom{0}and\phantom{0}} \varphi_t (x, 0) = \varphi^b_t (x, 0) ,
\end{equation}
where ${ \varphi^b (x, t) }$ denotes the stationary breather solution of the unperturbed SG equation in the infinite line, i.e.,
\begin{equation}
\label{eqn:6}
\varphi^b (x, t) = 4 \arctan \left[ \frac{\sqrt{1 - \omega^2}}{\omega} \frac{ \sin \left( \omega t + \vartheta \right) } { \cosh \left( \sqrt{1 - \omega^2} x \right) } \right] .
\end{equation}
Here, ${ 0 < \omega < 1 }$ and ${ 0 \leq \vartheta \leq 2 \pi }$ are, respectively, the frequency and the phase of the excitation, which is chosen with null velocity to only deal with its internal degree of freedom first. Also, in each numerical realization, ${ \vartheta }$ is randomly extracted from the above interval to introduce a degree of uncertainty in the starting state.

So far, no connection to any specific application realm has been made. However, it may be worth elucidating the physical meaning behind the previous expressions, referring, e.g., to the case of an LJJ. Equations~\eqref{eqn:1}-\eqref{eqn:4} accurately model the electrodynamics of an overlap-geometry Josephson tunnel junction of length $ L $ in the presence of a randomly varying bias current~\cite{Kivshar_1989}, thermal fluctuations~\cite{Castellano_1996}, and zero external magnetic field. In particular, Eq.\eqref{eqn:1} is obtained by normalizing space to the Josephson penetration depth ${ \lambda_J }$ and time to the inverse of the Josephson plasma frequency ${ \omega_p }$~\cite{Barone_1982}, with ${ \varphi (x, t) }$ representing the phase difference between the superconducting order parameters, ${ \alpha }$ accounting for the tunneling of quasiparticles, ${ \Gamma }$ and ${ \Gamma_T \propto T }$ being the noise amplitudes ($ T $ -- absolute temperature of the device), and ${ l = L / \lambda_J }$ giving the normalized junction's length. Furthermore, as stated in Sec.~\ref{sec1}, Eqs.~\eqref{eqn:5}-\eqref{eqn:6} are equivalent to assuming that a breather at a certain frequency is successfully excited at the junction's center at ${ t = 0 }$.

As for the computational details, Eq.~\eqref{eqn:1} is integrated by means of an implicit finite-difference scheme~\cite{Ames_1977}, with the stochastic terms approximated according to Refs.~\cite{Garcia_2012, Tuckwell_2016}. The discretization steps are $ \Delta x = \Delta t = 0.005 $, the system length is ${ l = 50 }$, and the observation time is fixed to ${ t_{\rm{obs}} = 250 }$.

\section{Results and discussion}
\label{sec3}

The results obtained for an increasingly complex perturbed SG equation are presented here. Specifically, Sec.~\ref{sec3a} gives an overview of the (noise-free) dissipative case, while Sec.~\ref{sec3b} and Sec.~\ref{sec3c} are concerned with, respectively, the effects of the spatially-homogeneous noisy force and the robustness of the findings with respect to the additional thermal background.

\subsection{Prelude: dissipative breather dynamics}
\label{sec3a}

Since both Sec.~\ref{sec3b} and Sec.~\ref{sec3c} require understanding of the purely dissipative case, i.e., Eq.~\eqref{eqn:1} with ${ \gamma(t) = \gamma_{T}(x, t) = 0 }$, the latter is covered here.

\begin{figure*}[t!!]
\centering
\includegraphics[width=\textwidth]{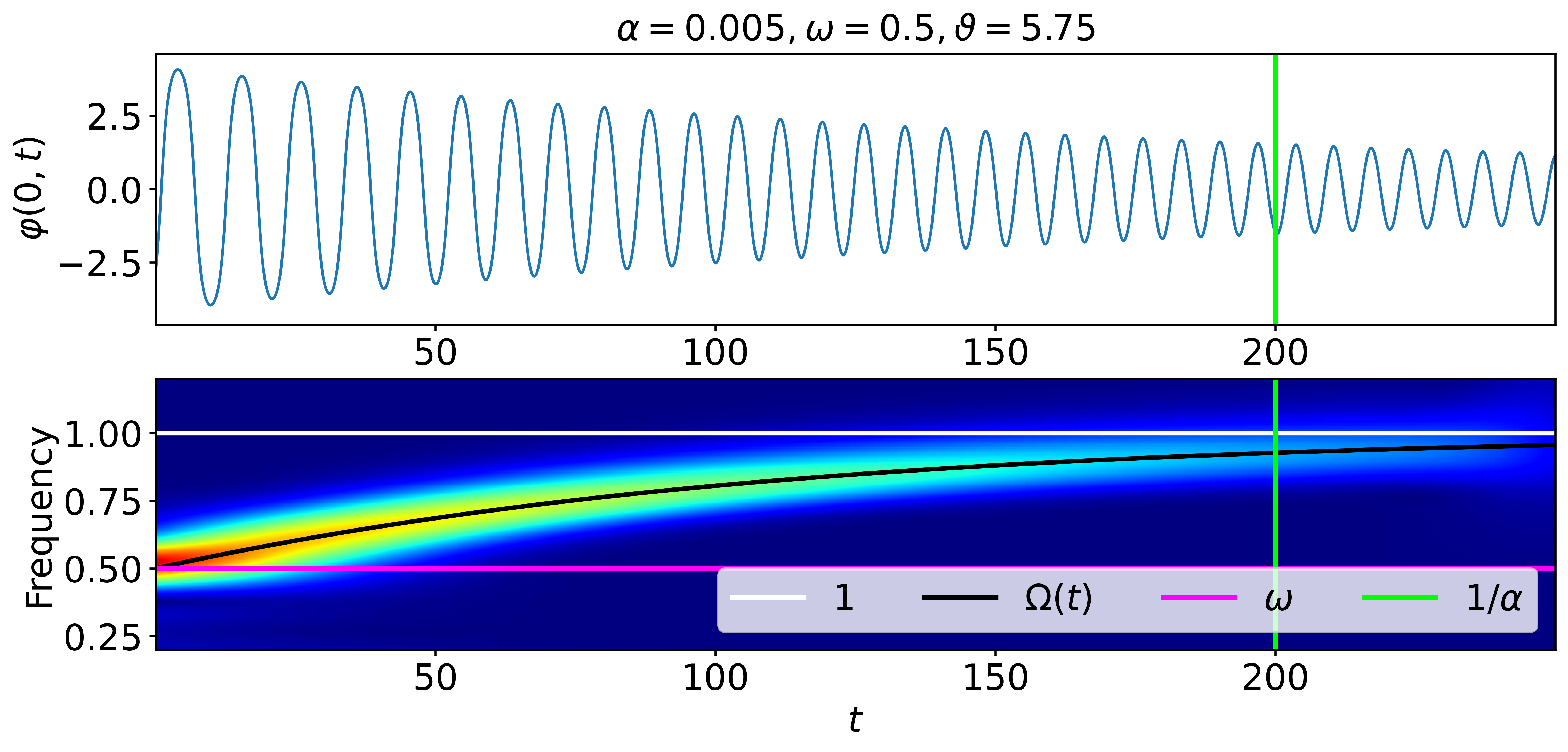}
\caption{\textit{Upper panel:} Amplitude of the decaying breather as a function of time. \textit{Lower panel:} Scalogram of the signal in the upper panel, i.e., the absolute value of its continuous wavelet transform plotted versus time and frequency. The color scale goes from the minimum (blue) to the maximum (red) value. The white and fuchsia horizontal lines correspond, respectively, to the plasma and the initial frequency, while the black curve indicates the estimated dissipation-induced modulation of the breather frequency, i.e., Eq.~\eqref{eqn:7}, and the green vertical line denotes the instant ${ t = 1 / \alpha }$. Simulation parameters: ${ l = 50 }$, ${ t_{\rm{obs}} = 250 }$, ${ \alpha = 0.005 }$, ${ \omega = 0.5 }$, and ${ \vartheta = 5.75 }$.}
\label{fig:1}
\end{figure*}
As discussed in previous literature~\cite{McLaughlin_1978_PRA, Gulevich_2006}, due to losses a breather decays into radiation within a time interval ${ \sim 1 / \alpha }$. Such a phenomenon is shown in Fig.~\ref{fig:1} for ${ \alpha = 0.005 }$, ${ \omega = 0.5 }$, and ${ \vartheta = 5.75 }$. More precisely, since Eqs.~\eqref{eqn:5}-\eqref{eqn:6} are used, the evanescent breather can be tracked by simply recording the field's value at ${ x = 0 }$, i.e., ${ \varphi (0, t) }$ (Fig.~\ref{fig:1}, upper panel). In fact, as long as no perturbing term explicitly depends on $ x $, the excitation cannot move away from its original position. By looking at the decaying amplitude profile, that is accurately captured by an exponential curve, a time-modulated frequency of oscillation may be noticed. In this regard, a powerful analysis tool is the scalogram~\cite{scaleogram} of the signal ${ \varphi (0, t) }$, i.e., the absolute value of its continuous wavelet transform~\footnote[1]{In the paper, the Complex Morlet wavelet function is employed.} versus time and frequency, which clearly shows (Fig.~\ref{fig:1}, lower panel) that the mode's frequency is evolving towards that of plasma waves (i.e., the white horizontal line). Moreover, by allowing for a time-dependent frequency ${ \Omega (t) }$ in the breather solution, i.e., in Eq.~\eqref{eqn:6}, and imposing a perfectly exponential trend for its amplitude, one finds the analytical expression
\begin{equation}
\label{eqn:7}
\Omega (t) = \cos \left[ \left( \arccos \omega \right) \exp \left( - \alpha t \right) \right] ,
\end{equation}
which excellently agrees with the numerical calculations (Fig.~\ref{fig:1}, lower panel, black curve). Equation~\eqref{eqn:7} also follows from the celebrated McLaughlin--Scott perturbation theory~\cite{McLaughlin_1978_PRA}, and it works well regardless of the specific ${ \alpha }$ and ${ \omega }$ combination. Indeed, the parameter values ${ \alpha = 0.005 }$ and ${ \omega = 0.5 }$ were chosen for visualization purposes only~\footnote[2]{Namely, since the time modulation of the breather frequency is gently spread over the entire simulation, the scalogram in Fig.~\ref{fig:1} is easier to read.}, whereas ${ \vartheta = 5.75 }$ was randomly extracted.

\subsection{Effects of the spatially-homogeneous noisy force}
\label{sec3b}

As the intensity ${ \Gamma }$ of the incoherent term ${ \gamma(t) }$ is increased (while keeping ${ \Gamma_{T} = 0 }$ for the moment), see Eqs.~\eqref{eqn:1}-\eqref{eqn:2}, the overall dynamical picture becomes progressively more complicated than that presented in Sec.~\ref{sec3a}. The radiative decay into a spatially-uniform state is no longer the only possible outcome. For example, the breather can gain enough energy for the splitting into a kink-antikink pair to occur, or the attenuation of the oscillations of the field ${ \varphi(x, t) }$ can be observed around ${ x = 0 }$ along with the emergence of solitonic patterns elsewhere [recall that ${ \gamma(t) }$ acts on the whole spatial domain, therefore it possibly amplifies plasma-like excitations even away from ${ x = 0 }$]. Therefore, the discrete set of oscillation peaks of ${ | \varphi (0, t) | }$ [indicated below by ${ \varphi_{\rm{max}} (t) }$], which tracks the time-varying breather amplitude, is not necessarily characterized by a monotonically decreasing behavior, as it would happen for ${ \gamma(t) = 0 }$, see the upper panel of Fig.~\ref{fig:1}. In light of this, by looking at ${ | \varphi (0, t) | }$ and ${ \varphi_{\rm{max}} (t) }$, a sort of first-passage time naturally comes to mind as a simple, but insightful, quantity to calculate for the analysis of the breather mode's evolution, with the initial state serving as an important reference point. To be more specific, substantial differences can arise from the initial breather configuration mainly due to the noisy perturbation driving the system towards a kink-antikink regime, or due to radiative losses. To deal with these two opposite cases, the following time-scales can be introduced, respectively 
\begin{equation}
\label{eqn:8}
\tau_{2 \pi} = \begin{cases}
\text{lowest\phantom{0}} \bar{t} \leq t_{\rm{obs}} \text{\phantom{0}such that\phantom{0}} | \varphi \left( 0, \bar{t} \right) | = 2 \pi \\
t_{\rm{obs}} \text{\phantom{0}if\phantom{0}} | \varphi \left( 0, t \right) | < 2 \pi \text{\phantom{0}} \forall t \leq t_{\rm{obs}} 
\end{cases}
\end{equation}
and
\begin{equation}
\label{eqn:9}
\tau_{\frac{1}{2}} = \begin{cases}
\text{lowest\phantom{0}} \widetilde{t} \leq t_{\rm{obs}} \text{\phantom{0}such that\phantom{0}} \varphi_{\rm{max}} \left( \widetilde{t} \right) = \varphi^b_{\frac{1}{2}} \\
t_{\rm{obs}} \text{\phantom{0}if\phantom{0}} \varphi_{\rm{max}} \left( t \right) > \varphi^b_{\frac{1}{2}} \text{\phantom{0}} \forall t \leq t_{\rm{obs}} ,
\end{cases}
\end{equation}
with ${ \varphi^b_{\frac{1}{2}} = 2 \arctan \left( \frac{\sqrt{1 - \omega^2}}{\omega} \right) }$ being equal to half of the unperturbed amplitude [obtained by substituting ${ x = 0 }$ and ${ \sin \left( \omega t + \vartheta \right) = 1 }$ in Eq.~\eqref{eqn:6}]. A characteristic time $ \tau $ can then be defined as
\begin{figure*}[t!!]
\centering
\includegraphics[width=0.6\textwidth]{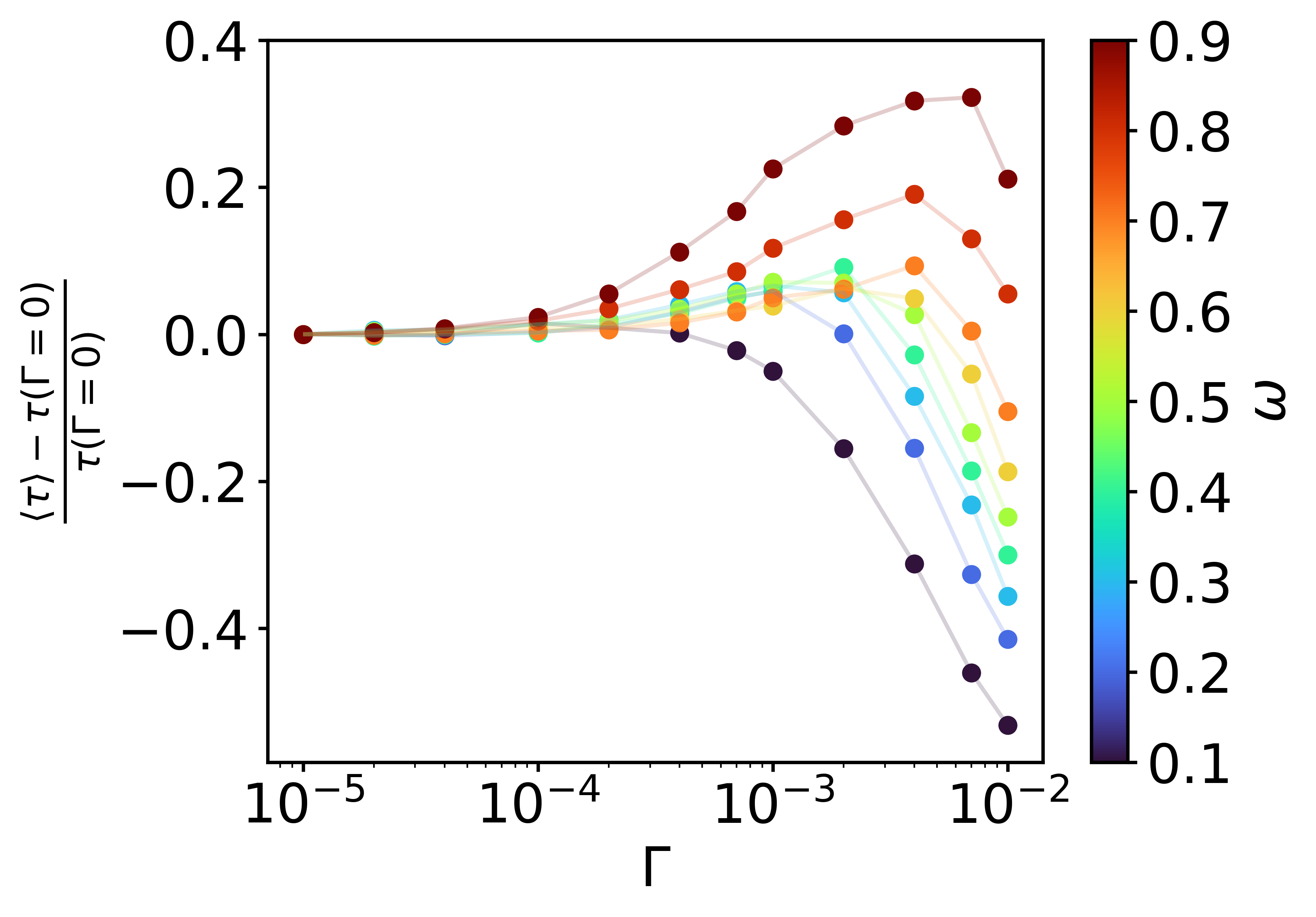}
\caption{Relative change of the average characteristic time ${ \left\langle \tau \right\rangle }$ as a function of the noise intensity ${ \Gamma \in \left[ 10^{-5} , 10^{-2} \right] }$, for different initial frequencies ${ \omega \in \left[ 0.1, 0.9 \right] }$, with ${ \Delta \omega = 0.1 }$. Other simulation parameters: ${ l = 50 }$, ${ t_{\rm{obs}} = 250 }$, ${ \alpha = 0.02 }$, and ${ N = 5 \times 10^3 }$.}
\label{fig:2}
\end{figure*}
\begin{equation}
\label{eqn:10}
\tau = \min \left( \tau_{2 \pi}, \tau_{\frac{1}{2}} \right) ,
\end{equation}
and, in a given simulation, it corresponds to the instant at which the phase profile has significantly departed from that of the initial excitation, whose persistence is the most relevant aspect. This represents a characteristic time for the breather, being a proxy of its lifetime in the presence of noise. Note that either successive kink-antikink recombinations and revivals of the field oscillations may stochastically take place for ${ t > \tau }$, but at the present stage these are regarded, in a sense, as higher-order phenomena.

Taking the deterministic value ${ \tau ( \Gamma = 0 ) \sim 1 / \alpha }$ as a reference, Fig.~\ref{fig:2} displays the relative change of the average characteristic time ${ \left\langle \tau \right\rangle }$, over $ N $ numerical runs, as a function of the noise intensity ${ \Gamma \in \left[ 10^{-5} , 10^{-2} \right] }$, for a set of initial frequencies ${ \omega \in \left[ 0.1, 0.9 \right] }$, with ${ \Delta \omega = 0.1 }$. As for the other simulation parameters, ${ \alpha = 0.02 }$~\footnote[3]{This is a typical value in the LJJ system~\cite{De_Santis_2022, De_Santis_2022_arxiv}. Furthermore, the choice of ${ \alpha }$ results in a much lower deterministic breather lifetime than that encountered in Sec.~\ref{sec3a}, thus it allows to address possible cases where ${ \tau \gg 1 / \alpha = 50 }$ without having to increase the observation time ${ t_{\rm{obs}} = 250 }$.} and ${ N = 5 \times 10^3 }$ are set hereafter. Moreover, as stated in Sec.~\ref{sec2}, each realization has an initial random phase ${ \vartheta \in [ 0, 2 \pi ] }$. A first remark concerning Fig.~\ref{fig:2} is that, independently of $ \omega $, there exists an extended range of $ \Gamma $ values where ${ \left\langle \tau \right\rangle }$ remains very close to its deterministic counterpart. In other words, a certain degree of robustness of the excitation against the stochastic force is observed. In addition, distinctively nonmonotonic behaviors emerge, i.e., the plot highlights a positive noise-induced effect on the breather stability, for almost all the considered initial frequencies.

\begin{figure*}[t!!]
\centering
\includegraphics[width=\textwidth]{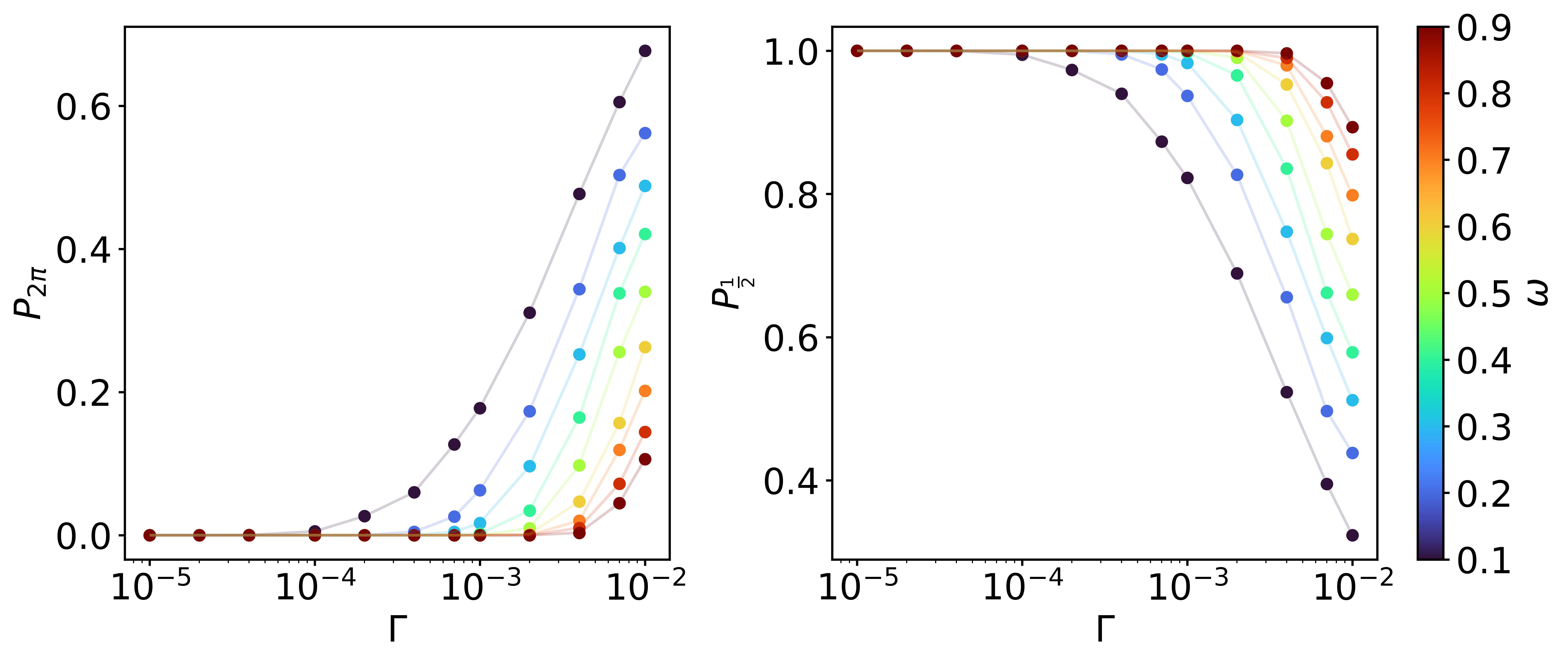}
\caption{Estimations of ${ P_{2 \pi} }$ (left panel) and ${ P_{\frac{1}{2}} }$ (right panel) versus the noise intensity ${ \Gamma \in \left[ 10^{-5} , 10^{-2} \right] }$, for different initial frequencies ${ \omega \in \left[ 0.1, 0.9 \right] }$, with ${ \Delta \omega = 0.1 }$. Other simulation parameters: ${ l = 50 }$, ${ t_{\rm{obs}} = 250 }$, ${ \alpha = 0.02 }$, and ${ N = 5 \times 10^3 }$.}
\label{fig:3}
\end{figure*}
\sloppy The distinct outcomes at the two ends of the breather's frequency spectrum can be understood by looking at the probabilities ${ P_{2 \pi} = \lim_{N \to \infty} N_{2 \pi} / N }$ and ${ P_{\frac{1}{2}} = \lim_{N \to \infty} N_{\frac{1}{2}} / N }$, with ${ N_{2 \pi} }$ being the number of cases where ${ \tau_{2 \pi} < \tau_{\frac{1}{2}} }$ and ${ N_{\frac{1}{2}} = N - N_{2 \pi} }$~\footnote[4]{In principle, both ${ \tau_{2 \pi} }$ and ${ \tau_{\frac{1}{2}} }$ could be equal to ${ t_{\rm{obs}} }$ for a certain realization, see Eqs.~\eqref{eqn:8}-\eqref{eqn:9}. However, taking ${ t_{\rm{obs}} \gg 1 /\alpha }$ essentially prevents this from happening, and besides one can adjust ${ N_{2 \pi} }$ and ${ N_{\frac{1}{2}} }$ accordingly.}, see Fig.~\ref{fig:3}. In short, as $ \Gamma $ is incremented, for higher initial amplitudes [yielding lower $ \omega $, see Eq.~\eqref{eqn:6}] the stochastic energy input from the ${ \gamma(t) }$ term is very likely to result in the splitting of the kink-antikink bond early in the simulations; instead, breathers with lower amplitudes (i.e., higher $ \omega $) can more easily be sustained, i.e., values of ${ \tau }$ significantly larger than ${ \tau( \Gamma = 0 ) }$ may be reached, without abandoning the energy range allowed for the oscillatory state.

With regard to the present discussion, another useful visualization tool can be obtained through the following time series
\begin{equation}
\label{eqn:11}
P (t, \Gamma) = \frac{1}{N} \sum_{n = 1}^{N} \rho_n (t, \Gamma) , \text{\phantom{0}where\phantom{0}} \rho_n (t, \Gamma) = \begin{cases}
1 \text{\phantom{0}if\phantom{0}} t \leq \tau_n (\Gamma) \\
0 \text{\phantom{0}if\phantom{0}} t > \tau_n (\Gamma) ,
\end{cases}
\end{equation}
${ n = 1, ..., N }$ goes over the different numerical runs, and ${ \tau_n (\Gamma) }$ is the characteristic time [see Eq.~\eqref{eqn:10}] relative to the $ n $-th realization performed with noise amplitude $ \Gamma $. For a given ${ \Gamma }$ value, a few basic features concerning ${ P (t, \Gamma ) }$ can be identified. First, ${ \int_0^\infty P(t, \Gamma) dt = \left\langle \tau \right\rangle \! (\Gamma) }$. Moreover, since for ${ \Gamma \to 0 }$ one has that ${ \tau_n (\Gamma) \sim 1 / \alpha }$ with very little difference among the repeated simulations, in this limit a steplike behavior is expected for ${ P (t, \Gamma ) }$.
\begin{figure*}[t!!]
\centering
\includegraphics[width=\textwidth]{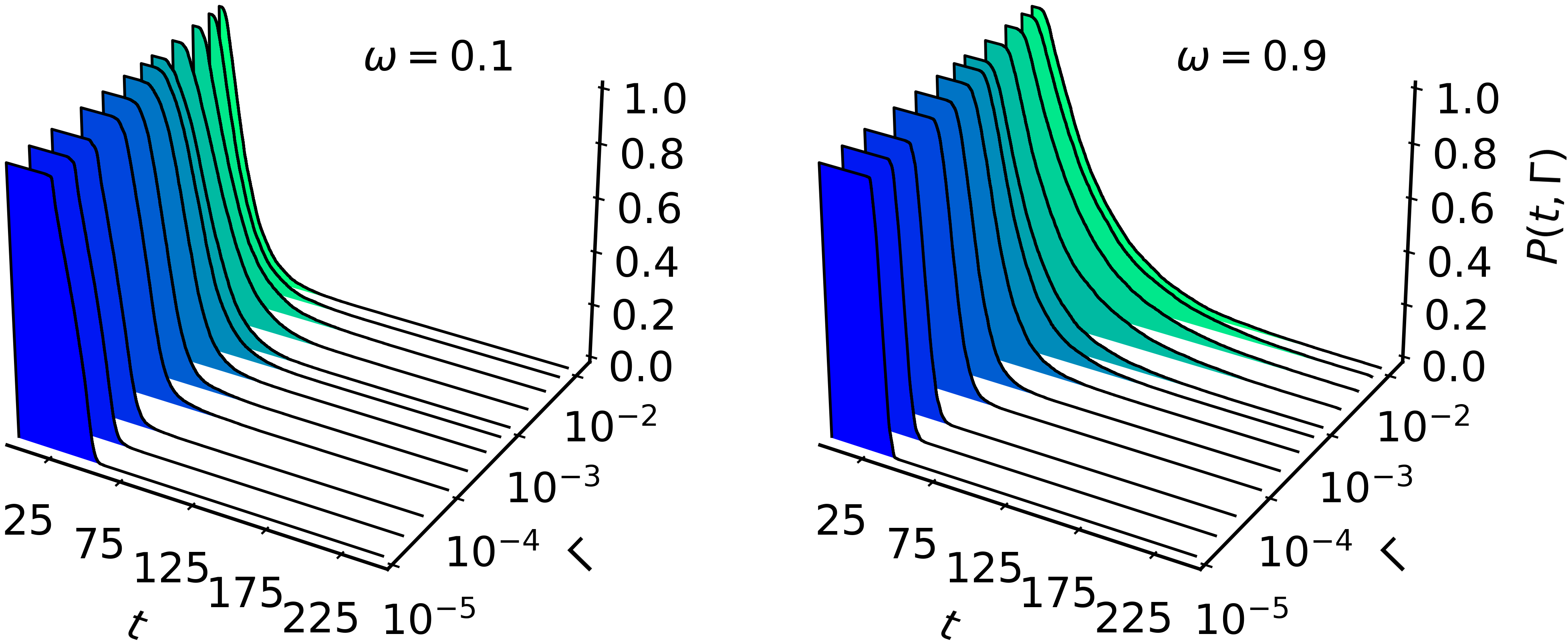}
\caption{Computations of the quantity ${ P (t, \Gamma) }$ [see Eq.~\eqref{eqn:11}], for ${ \omega = 0.1 }$ (left panel) and ${ \omega = 0.9 }$ (right panel). Other simulation parameters: ${ l = 50 }$, ${ t_{\rm{obs}} = 250 }$, ${ \alpha = 0.02 }$, and ${ N = 5 \times 10^3 }$.}
\label{fig:4}
\end{figure*}

Figure~\ref{fig:4} illustrates the computed ${ P (t, \Gamma) }$ profiles, for ${ \omega = 0.1 }$ (left panel) and ${ \omega = 0.9 }$ (right panel). Along with the above mentioned properties, the graph highlights the marked difference between the dynamics at opposite ends of the breather's frequency spectrum. In particular, for ${ \omega = 0.9 }$, the slower time-decay of ${ P (t, \Gamma ) }$ for ${ 10^{-3} < \Gamma < 10^{-2} }$ (i.e., in correspondence of the peak of ${ \left\langle \tau \right\rangle }$, see Fig.~\ref{fig:2}) indicates that the fluctuating force enables the system to reach values ${ \tau \gg 1 / \alpha = 50 }$.

\begin{figure*}[t!!]
\centering
\includegraphics[width=\textwidth]{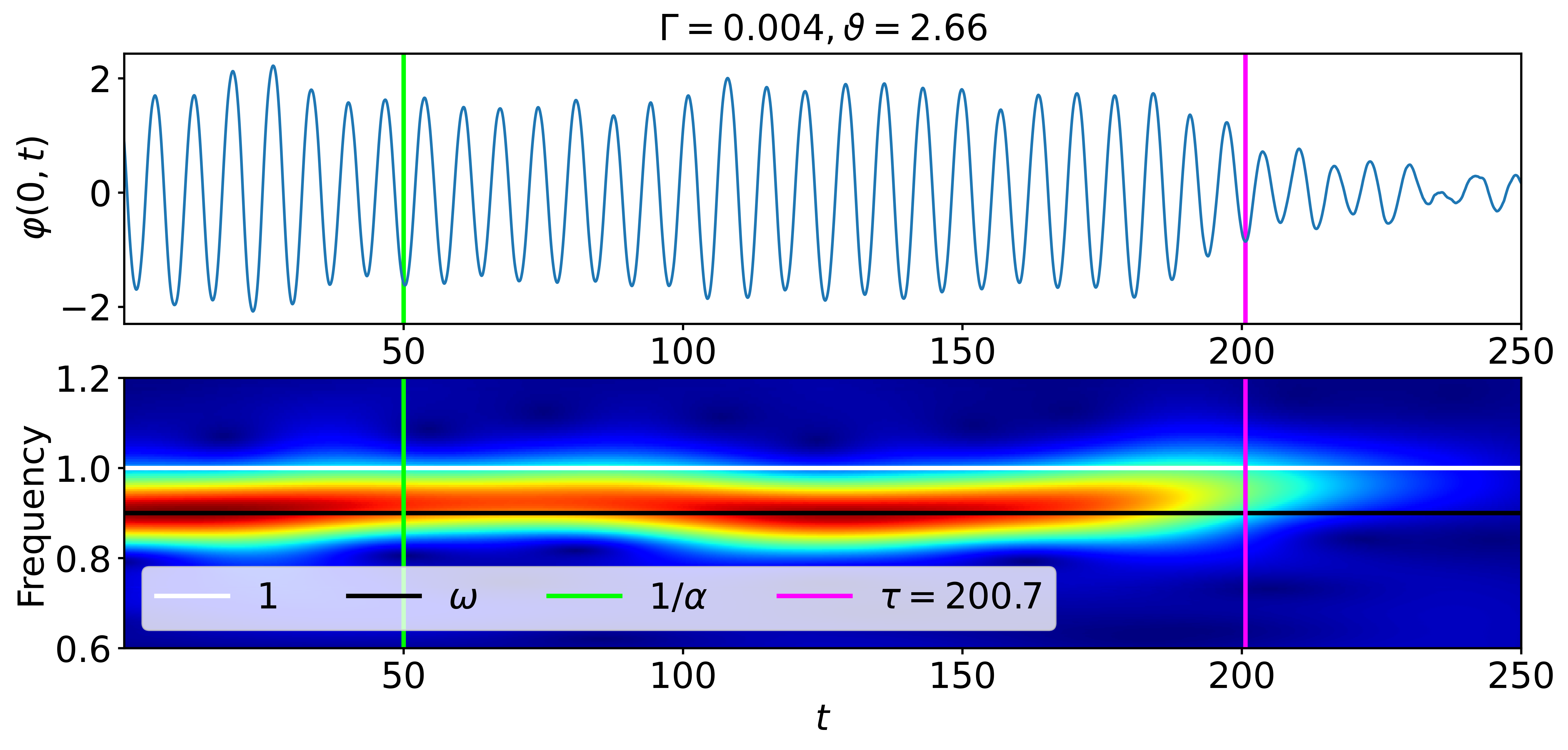}
\caption{\textit{Upper panel:} Breather amplitude as a function of time, for one of the realizations performed with the noise intensity ${ \Gamma = 4 \times 10^{-3} }$. \textit{Lower panel:} Scalogram of the signal in the upper panel, i.e., the absolute value of its continuous wavelet transform plotted versus time and frequency. The color scale goes from the minimum (blue) to the maximum (red) value. The white and black horizontal lines correspond, respectively, to the plasma and the initial frequency, while the green and fuchsia vertical lines indicate, respectively, the instants ${ t = 1 / \alpha }$ and ${ t = \tau = 200.7 }$. Other simulation parameters: ${ l = 50 }$, ${ t_{\rm{obs}} = 250 }$, ${ \alpha = 0.02 }$, ${ \omega = 0.9 }$, and ${ \vartheta = 2.66 }$.}
\label{fig:5}
\end{figure*}

The idea of a noise-enhanced stability effect on the breather state is quite intriguing, and should be supported by further evidence. Hence, the rest of the work is focused on ${ \omega = 0.9 }$, which is the more interesting case. To show that even for ${ t \gg 1 / \alpha }$ the breather's distinctive traits can be stochastically maintained, a close look is taken at one of the many realizations performed with the noise amplitude ${ \Gamma = 4 \times 10^{-3} }$ (i.e., approximately at the peak of ${ \left\langle \tau \right\rangle }$, see Fig.~\ref{fig:2}). In particular, the upper panel of Fig.~\ref{fig:5} shows that the oscillation amplitude remains very close to that of the unperturbed excitation roughly up to ${ t = \tau }$ (see the fuchsia vertical line), whereas the lower panel is concerned with the scalogram of ${ \varphi (0, t) }$. The latter plot is highly informative as well, since the persistence of $ \omega $ as the dominant frequency of the signal (see the black horizontal line) up to ${ t \approx \tau }$ provides an additional mean for documenting the effectiveness of the energy transfer from the noise signal to the nonlinear mode. Note also that, in contrast to Fig.~\ref{fig:1}, little or no attenuation of the peak value of the spectrogram is observed for the majority of the simulation.
\begin{figure*}[t!!]
\centering
\includegraphics[width=0.5\textwidth]{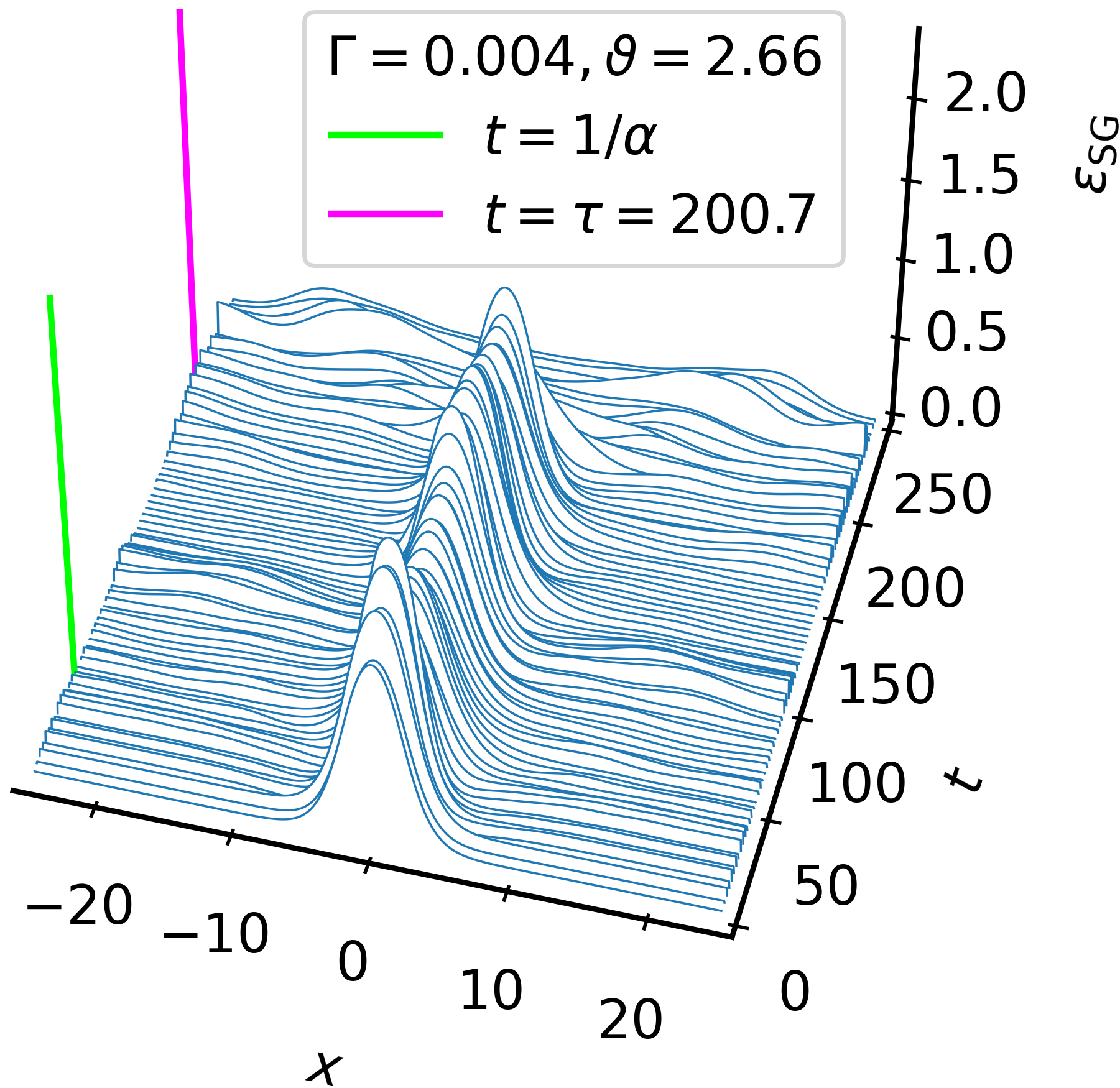}
\caption{Energy density ${ \varepsilon_{\rm{SG}} }$ [Eq.~\eqref{eqn:12}] as a function of space and time, for the same realization presented in Fig.~\ref{fig:5}. The green and fuchsia vertical lines indicate, respectively, the instants ${ t = 1 / \alpha }$ and ${ t = \tau = 200.7 }$.}
\label{fig:6}
\end{figure*}
The localization of energy in the spatial domain is another aspect to be examined when dealing with breather-like excitations. To this end, the energy density
\begin{equation}
\label{eqn:12}
\varepsilon_{\rm{SG}} = \frac{1}{2} \varphi_t^2 + \frac{1}{2} \varphi_x^2 + 1 - \cos \varphi 
\end{equation}
can be computed. For the same numerical run as before (see Fig.~\ref{fig:5}), Fig.~\ref{fig:6} evidently confirms that the strongly localized mode is indeed sustained far beyond its deterministic lifetime ${ 1 / \alpha = 50 }$ (see the green vertical line), approximately up to ${ t = \tau = 200.7 }$ (see the fuchsia vertical line).

\begin{figure*}[t!!]
\centering
\includegraphics[width=0.6\textwidth]{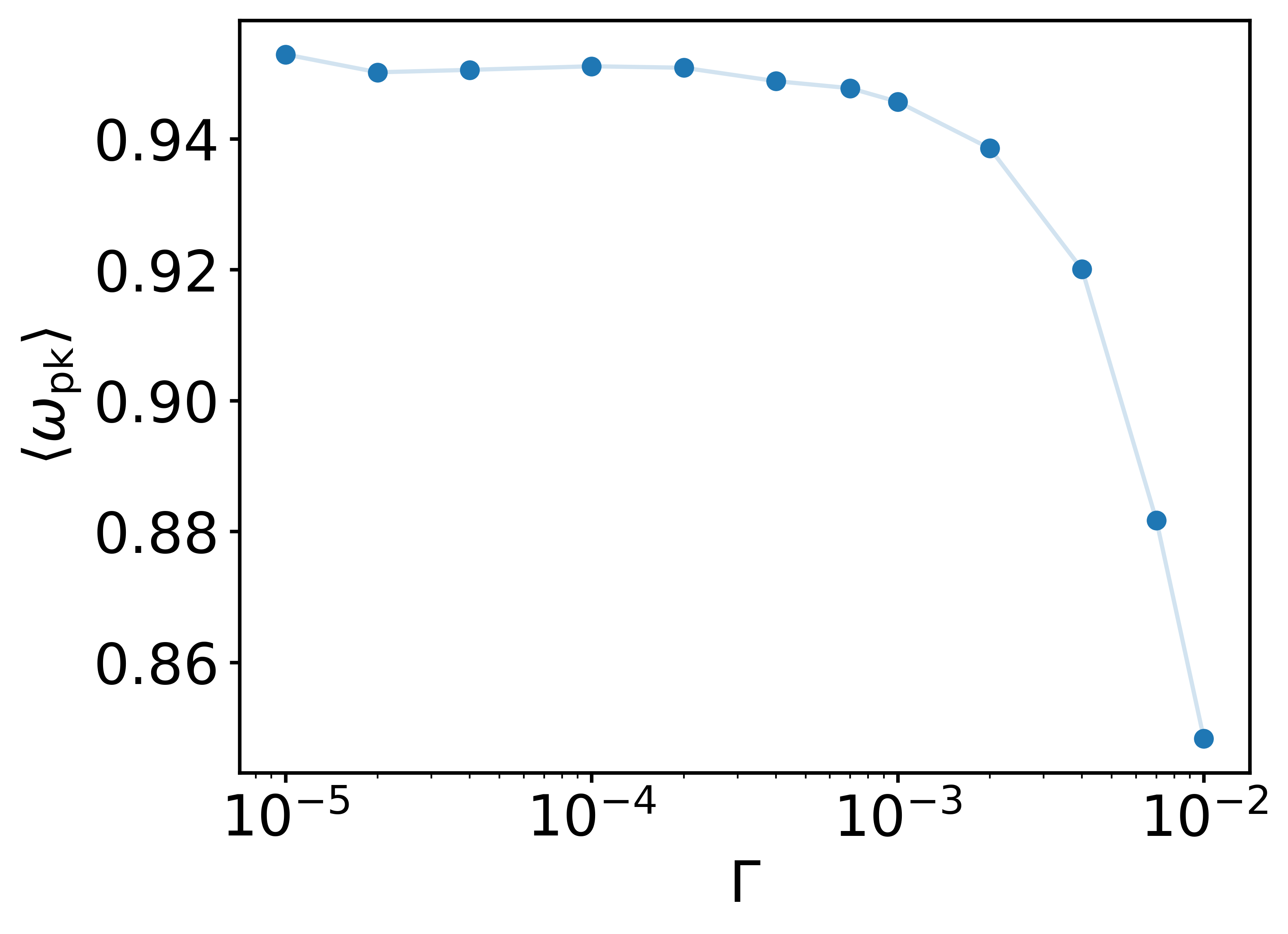}
\caption{Average peak frequency ${ \left\langle \omega_{\rm{pk}} \right\rangle }$ of the PSD of the truncated time series ${ \varphi ( 0, t \leq \tau ) }$ as a function of the noise intensity ${ \Gamma \in \left[ 10^{-5} , 10^{-2} \right] }$. Other simulation parameters: ${ l = 50 }$, ${ t_{\rm{obs}} = 250 }$, ${ \alpha = 0.02 }$, ${ N = 5 \times 10^3 }$, and ${ \omega = 0.9 }$.}
\label{fig:7}
\end{figure*}

Although the example covered in Figs.~\ref{fig:5}~and~\ref{fig:6} illustrates the phenomenon very clearly, variegated spatio-temporal patterns can arise due to the stochastic force, as stated at the beginning of this subsection. Therefore, it seems appropriate to look for an additional overall characterization of the noise-driven tendency towards the nonlinear oscillatory regime. A simple approach consists in (i)~evaluating the power spectral density (PSD) of the truncated time series ${ \varphi ( 0, t \leq \tau ) }$, (ii)~identifying the PSD's peak frequency ${ \omega_{\rm{pk}} }$ for each numerical run, and then (iii)~averaging ${ \omega_{\rm{pk}} }$ over the performed realizations. The quantity ${ \omega_{\rm{pk}} }$ presumably represents the most prominent frequency component of the signal of interest, and, in the case of a surviving nonlinear excitation, it is expected to differ from $ 1 $~\footnote[5]{Recall that this is the plasma value, i.e., the fundamental frequency of linear SG excitations.} more than in a Fig.~\ref{fig:1}-like scenario. The average peak frequency ${ \left\langle \omega_{\rm{pk}} \right\rangle }$ is shown as a function of $ \Gamma $ in Fig.~\ref{fig:7}. Interestingly, the noisy force drives the system away from the plasma frequency, towards lower ${ \left\langle \omega_{\rm{pk}} \right\rangle }$ values, which is a signature of the breather's persistence.

To conclude, it is worth adding that extensive tests have been conducted for the system lengths ${ l = 100, 200 }$ as well, and no modifications to the overall picture presented here occur.

\subsection{Robustness of the phenomenon against the additional thermal background}
\label{sec3c}

The complete version of Eq.~\eqref{eqn:1} is now examined for different values of ${ \Gamma_{T} }$. As mentioned in Sec.~\ref{sec3b}, the following simulation parameters are chosen: ${ \alpha = 0.02 }$, ${ \Gamma \in \left[ 10^{-5} , 10^{-2} \right] }$, ${ N = 5 \times 10^3 }$, and ${ \omega = 0.9 }$. For each realization, the phase ${ \vartheta }$ is randomly extracted from the interval ${ [ 0, 2 \pi ] }$ also in this subsection.

In contrast to the previous cases, due to the ${ \gamma_{T}(x, t) }$ term, the position of the breather's center fluctuates around ${ x = 0 }$. This can be accounted for while keeping the same definitions for ${ \tau }$, ${ \tau_{2 \pi} }$, and ${ \tau_{\frac{1}{2}} }$ [i.e., Eqs.~\eqref{eqn:8}-\eqref{eqn:10}]. Since exclusively looking at the phase profile at ${ x = 0 }$ is no longer very reasonable, one can consider a sampling region ${ -\delta/2 \leq x \leq \delta/2 }$, with ${ 0 < \delta \ll l }$, and work with the quantity
\begin{equation}
\label{eqn:13}
\max_{-\delta/2 \leq x \leq \delta/2} | \varphi (x, t) |
\end{equation}
instead of ${ | \varphi (0, t) | }$. Note that here ${ t = \tau_{\frac{1}{2}} }$ can potentially be reached in a given simulation due to the excitation's departure from the monitored region. The latter circumstance, however, does not affect the relevance of $ \tau $ as a characteristic time for the system. In fact, within a hypothetical experimental framework, the breather's escape from a given confined area would correspond to a breakdown of the conditions optimal for carrying out the measurements.

\begin{figure*}[t!!]
\centering
\includegraphics[width=\textwidth]{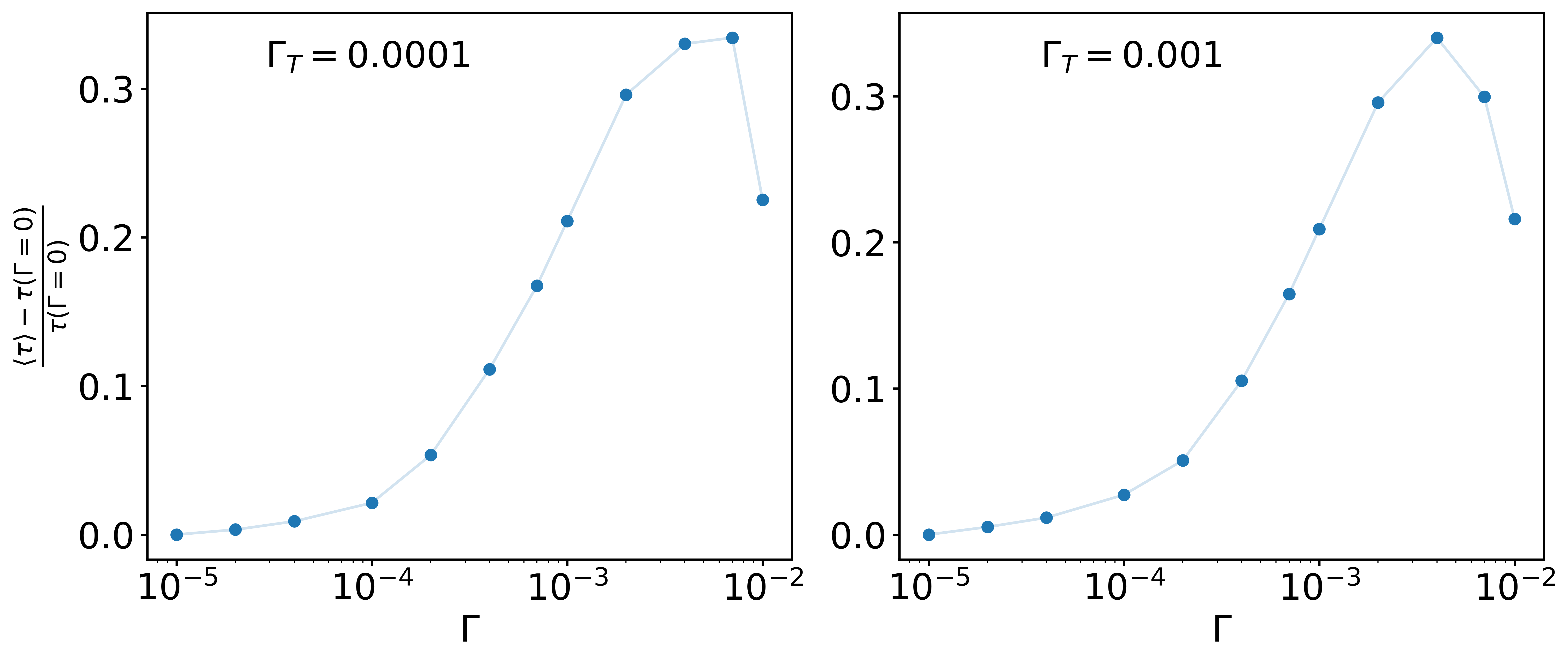}
\caption{Relative change of the average characteristic time ${ \left\langle \tau \right\rangle }$ as a function of the noise intensity ${ \Gamma \in \left[ 10^{-5} , 10^{-2} \right] }$, for ${ \Gamma_{T} = 10^{-4} }$ (left panel) and ${ \Gamma_{T} = 10^{-3} }$ (right panel). The length of the sampling region is ${ \delta = 5 }$. Other simulation parameters: ${ l = 50 }$, ${ t_{\rm{obs}} = 250 }$, ${ \alpha = 0.02 }$, ${ N = 5 \times 10^3 }$, and ${ \omega = 0.9 }$.}
\label{fig:8}
\end{figure*}
In a fashion similar to Fig.~\ref{fig:2}, the value ${ \tau ( \Gamma = 0 ) \sim 1 / \alpha }$ is taken as a reference in Fig.~\ref{fig:8}. Here, the relative change of ${ \left\langle \tau \right\rangle }$ is shown as a function of the noise intensity $ \Gamma $, for ${ \Gamma_{T} = 10^{-4} }$ (left panel) and ${ \Gamma_{T} = 10^{-3} }$ (right panel). Distinct nonmonotonic behaviors are seen in both cases. Furthermore, the two profiles are very close to each other, and to that of Fig.~\ref{fig:2} for ${ \omega = 0.9 }$ also, which is a good indication of the result's robustness. The length of the sampling region is ${ \delta = 5 }$, however several values ${ \delta \lesssim \lambda_b }$ have been tested, where ${ \lambda_b }$ represents the typical spatial extension of the breather, without substantial alteration of the presented curves.

\section{Conclusions}
\label{sec4}

This work numerically explores the influence of both dissipative and stochastic perturbing terms on the dynamics of SG breathers. Starting from a stationary breather with a random phase value, it is shown that a spatially-homogeneous noisy force can make the oscillating state last far beyond its well-known radiative decay lifetime~\cite{McLaughlin_1978_PRA, Gulevich_2006}. The documented noise-enhanced stability phenomenon appears to be most effective on the upper side of the breather frequency spectrum, where a more pronounced nonmonotonic behavior of an average characteristic time is observed as a function of the noise intensity. The existence of a noise-driven trend towards the nonlinear mode is also confirmed by analyzing the frequency domain and the energy localization. The result appears to be robust against an additional thermal background.

The present framework can, in principle, be used in experiments for the detection of breathers in LJJs~\cite{Gulevich_2012} and in their subsequent applications~\cite{Macias-Diaz_2007, Macias-Diaz_2007_1, Fujii_2007, Fujii_2008, Fuj09}. Furthermore, considering the degree of universality of the SG model, detailed investigations that provide insight into its fundamental excitations possess an inherent interdisciplinary scope. In this regard, SG breathers have recently attracted research interest, e.g., in cuprate superconductors~\cite{Dienst_2013}, geology~\cite{Zalohar_2020}, and DNA systems~\cite{Liu_2021}.

Finally, the topic of stochastic SG breather dynamics is far from being fully understood. It may be worth investigating the role of fluctuations with different statistical properties and/or spatial modulations. For instance, one could examine a colored noise source with cutoff frequency $ \omega_{\rm{cutoff}} $ and establish whether varying the difference ${ \omega_{\rm{cutoff}} - \omega }$ (with $ \omega $ being the internal breather frequency) produces any significant change in the response of the solitonic mode.

\section*{Acknowledgments}
\label{acknowledgments}

This work was supported by the Government of the Russian Federation through Agreement No. 074-02-2018-330 (2) and partially by the Italian Ministry of University and Research (MUR).


\end{document}